\documentclass[
    ,final            
  ]
  {aipproc}

\layoutstyle{8x11single}

\usepackage{amssymb,graphicx}
\newcommand\lsim{\mathrel{\rlap{\lower4pt\hbox{\hskip1pt$\sim$}}
    \raise1pt\hbox{$<$}}}
\newcommand\gsim{\mathrel{\rlap{\lower4pt\hbox{\hskip1pt$\sim$}}
    \raise1pt\hbox{$>$}}}

\newcommand{\captionfonts}{\small}

\makeatletter  
\long\def\@makecaption#1#2{%
  \vskip\abovecaptionskip
  \sbox\@tempboxa{{\captionfonts #1: #2}}%
  \ifdim \wd\@tempboxa >\hsize
    {\captionfonts #1: #2\par}
  \else
    \hbox to\hsize{\hfil\box\@tempboxa\hfil}%
  \fi
  \vskip\belowcaptionskip}
\makeatother   

\begin{document}
\title{Interplay of Direct and Indirect Searches for New Physics\footnote{Based on an invited talk given by TH at the 19th Particles and Nuclei International Conference  PANIC11, MIT, Cambridge, USA, 24th-29th July 2011.  MZ-TH/11-28} }


\author{Tobias Hurth}{
  address={Institute for Physics (THEP), Johannes Gutenberg University Mainz\\
  D-55099 Mainz, Germany}
}

\author{Sabine Kraml}{
  address={Laboratoire de Physique Subatomique et de Cosmologie, UJF Genoble 1,\\
  CNRS/IN2P3, INPG, F-38026 Grenoble, France}
}

\begin{abstract}
 We report recent work on the interplay of collider and flavour physics 
 regarding the search for physics beyond the Standard Model.
\end{abstract}

\maketitle


\section{Introduction}
At the beginning of the LHC era, the search for new degrees of freedom 
beyond the Standard Model (SM) is within the main focus  of particle 
physics. In principle, there are two ways to search for possible new 
degrees of freedom. 
At the high-energy frontier one tries to produce and observe them directly, 
while at the high-precision frontier one analyses their indirect virtual 
effects within flavour or electroweak observables.

Flavour changing neutral currents (FCNCs) test the SM  at  the one-loop level
and offer complementary information about the SM and it extensions. 
They are sensitive for the product of mixing angles and mass splittings 
between possible new heavy particles while within the direct search 
the masses can be measured directly. 

Within supersymmetric extensions of the SM, the measurement of the flavour structure 
is directly linked to the crucial question of the supersymmetry-breaking mechanism. 
Thus, the  flavour sector is important in distinguishing between models of 
supersymmetry. This example demonstrates the obvious complementary nature of 
flavour physics and high-$p_T$  physics.  
At the LHC, direct searches for supersymmetric particles are essential for 
establishing the existence and the nature of new physics (NP) beyond the SM. 
On the other hand, flavour physics provides an important tool with which 
fundamental questions regarding the structure of this NP, 
such as how supersymmetry is broken, can be addressed.

None of the dedicated flavour experiments in the last decade has observed any 
unambiguous sign of new physics yet~\cite{Antonelli:2009ws}, 
in particular there are no ${\cal O}(1)$ NP effects in any FCNC 
process~\cite{Hurth:2010tk}. 
Also the first results of the LHCb experiment~\cite{LHCb}
are in full agreement with the CKM theory of the SM.
This experimental fact implies the infamous flavour problem of NP, 
namely why FCNC processes are suppressed. 
It has to be solved in any viable NP model. The hypothesis of minimal flavour 
violation (MFV)~\cite{D'Ambrosio:2002ex}, i.e.\ that the NP model
has no flavour structures beyond the Yukawa couplings, solves this problem formally. 
A completely anarchic flavour structure, on the other hand, would require that 
the scale of NP be tens of TeV, such that NP effects decouple~\cite{Isidori:2010kg}. 
In between these two extremes, non-minimal flavour violation can still be compatible  
with the present data (see below) because the flavour sector has been tested only 
at the $10\%$ level in $b\to s$  transitions. 

From theoretical arguments we expect NP to become 
apparent at the ${\cal O}(1)$~TeV scale, while flavour constraints naturally seem to point to a much 
higher scale. In the case of supersymmetry in particular,  flavour constraints limit sparticles 
from being very light.   On the other hand, 
ATLAS and CMS searches with about 1~fb$^{-1}$ of data at 7~TeV~\cite{lhc7} already 
put lower limits on squark and gluino masses of roughly $m_{\tilde q,\tilde g}\gtrsim1.1$~TeV 
for $m_{\tilde q}\simeq m_{\tilde g}$, thus beginning to probe the 
preferred regions of simple SUSY realizations like the CMSSM. 
It is worth noting, however, that the limits on the gluino mass from the current LHC run 
become much weaker when $1^{st}/2^{nd}$ generation squarks are somewhat 
heavier---as indeed preferred by the flavour constraints. Sub-TeV stops 
and electroweak  gauginos  are still  allowed.

The CERN working group 
``Interplay of Collider and Flavour Physics''~\cite{Twiki} 
addresses the complementarity and synergy between the LHC and the flavour factories 
within the new physics search. It is a follow-up of the two recent CERN workshop 
series ``Flavour in the Era of the LHC''~\cite{LHCFlavour} and  
``CP Studies and Non-Standard Higgs Physics''~\cite{CPHiggs} 
at the interface of collider and flavour physics and experiment and theory.

In this contribution, we report on some recent work  on this interplay. 
For lack of space we focus on two examples, and we apologize for the 
omission of other relevant work.

\section{Flavour-violating squark decays} 

Flavour-violating high- and low-energy observables are governed by the
same parameters in supersymmetric models. A particularly  important
question is whether the soft SUSY breaking parameters can have
additional flavour structures beyond the well-known CKM.
Recently, it was shown \cite{Hurth:2009ke} that,  
in view of the present flavour data, flavour-violating squark 
and gluino decays can be typically of order  $10\%$ in the regions of
parameter space where no or only moderate cancellations between
different contributions to the low energy observables occur~\footnote{If one allows for
larger  new physics contributions, e.g.~the same order as the SM
contributions, in the flavour observables, then even
flavour-violating branching ratios of up to $40\%$ are consistent with
the present data.}.  
This observation has an impact on the discovery
strategy of squarks and gluinos as well as on the measurement of the underlying
parameters at the LHC.  For example, in mSUGRA points without flavour mixing one
 finds usually that
the left-squarks of the first two generations as well as the right squarks
have similar masses. However, large flavour mixing implies that there is a considerable
mass splitting as can be seen. Therefore, the assumption
of almost degenerate masses should be reconsidered and the possibility of 
sizeable flavour-changing squark and gluino decays taken into account.

\begin{figure}[t]
\includegraphics[height=3.0cm,width=6cm]{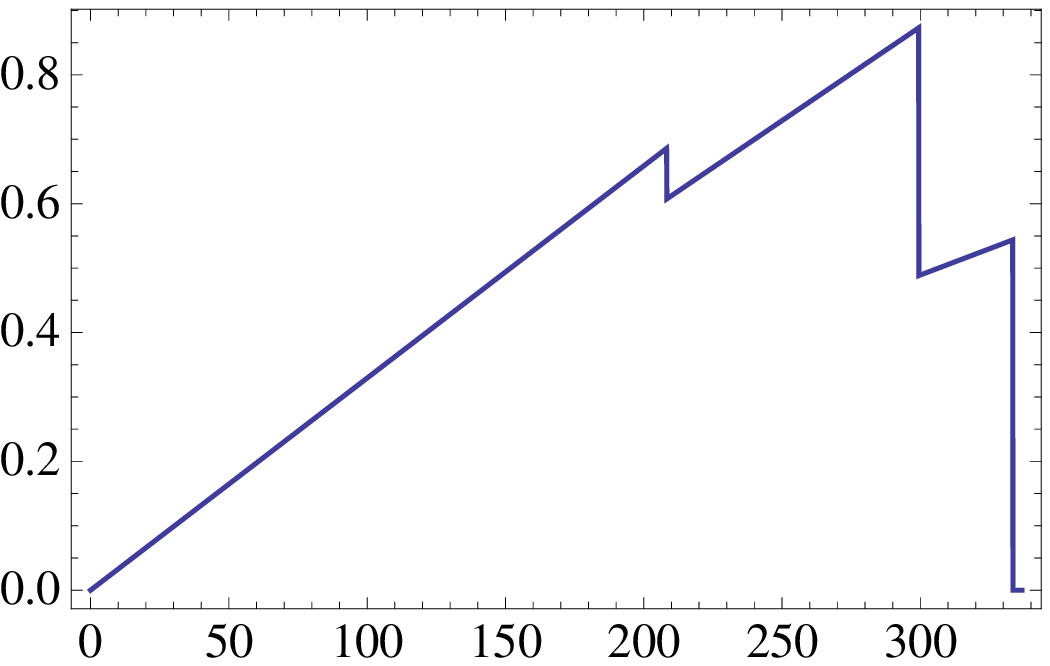} $m_{bb}$\qquad 
\includegraphics[height=3.0cm,width=6cm]{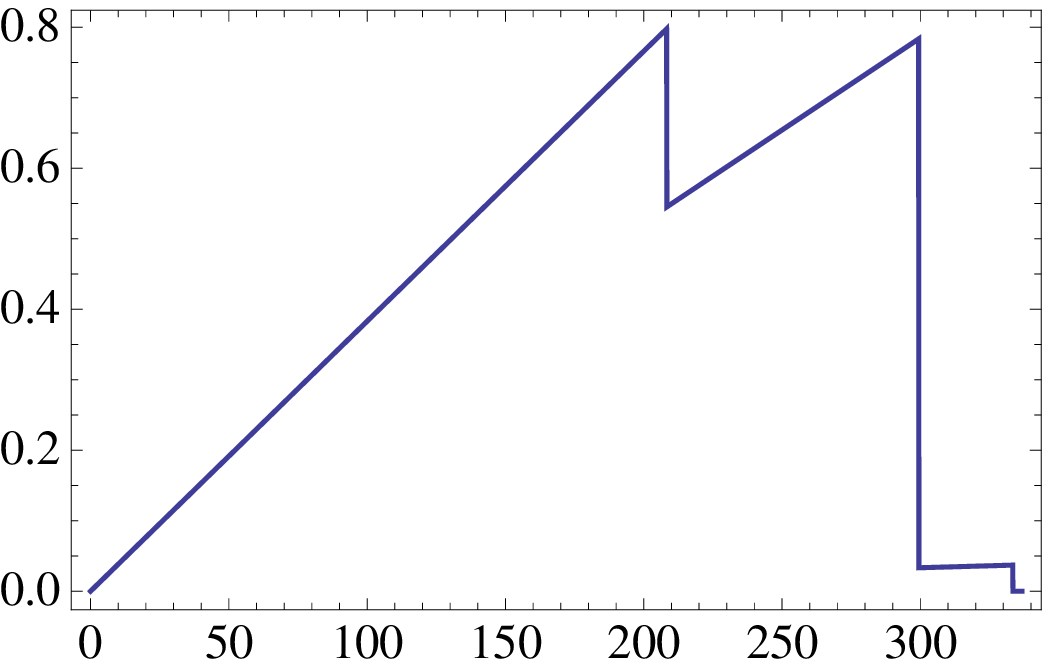} $m_{bs}$
\caption{Left: Differential distributions $d$BR($\tilde g \to b \bar{b} \tilde \chi^0_1)/d m_{bb}\times 10^4$
as a function of $m_{bb} = \sqrt{(p_b + p_{\bar{b}})^2}$.
Right:
$d$(BR($\tilde g \to b s  \tilde \chi^0_1)/d m_{bs}$  as a function of  $m_{bs}$  (the sum over the charges is shown:
BR($\tilde g \to b \bar{s} \tilde \chi^0_1)$ + BR$(\tilde g \to \bar{b} s \tilde \chi^0_1)$).
 }
\label{fig:bbbar}
\end{figure}

An important part of the decay chains considered for SPS1a' and nearby points
are $\tilde g \to b \tilde b_j \to b \bar{b} \tilde \chi^0_k$ which are
used to determine the gluino mass as well as the sbottom masses or at least
their average value if these masses are close \cite{Branson:2001ak}. 
In the latter analysis, the existence of two $b$-jets has been assumed
stemming from this decay chain. In this case the two contributing
sbottoms would lead to two edges in the partial distribution
$d$(BR($\tilde g \to b \bar{b} \tilde \chi^0_1)/d m_{bb}$ where
$m_{bb}$ is the invariant mass of the two bottom quarks.  As can be
seen from the left plot of Figure~\ref{fig:bbbar}, there are scenarios where more
squarks can contribute and consequently one finds a richer structure,
e.g.~three edges in the example shown. 
Such a structure is either a clear sign of flavour violation or the
fact that the particle content of the MSSM needs to be extended. 
The edge analysis allows both options.  
The differential distribution of the final state $b s\tilde \chi^0_1$ 
shows a similar structure where the edges occur at
the same places as in the $b\bar{b}$ spectrum but with different
relative heights, see right plot of Figure~\ref{fig:bbbar}. 
This gives a non-trivial cross-check on the
hypothesis of sizeable flavour mixing. Clearly a detailed Monte Carlo
study will be necessary to see with which precision one can extract
information on these edges.

\section{Lepton-flavour violation in 5D SUSY-GUTs} 

In supersymmetric models with a GUT-sized extra dimension, 
where both the Higgs fields and the SUSY breaking hidden sector 
are localized on a 4D brane, exponential wave function profiles of 
the matter fields give rise to hierarchical structures in the Yukawa 
couplings and soft terms. Such structures can naturally explain 
hierarchical fermion masses and mixings, while at the same time 
alleviating the supersymmetric flavour problem.
This idea and its phenonmenological consequences have been discussed 
in the literature mostly on the qualitative level, see e.g.~\cite{sedim}. 
Very recently a detailed numerical study for this class of models 
was performed in \cite{Brummer:2011cp} for two sources of supersymmetry 
breaking, radion mediation and brane fields. 
  
\begin{figure}
\includegraphics[height=5.5cm]{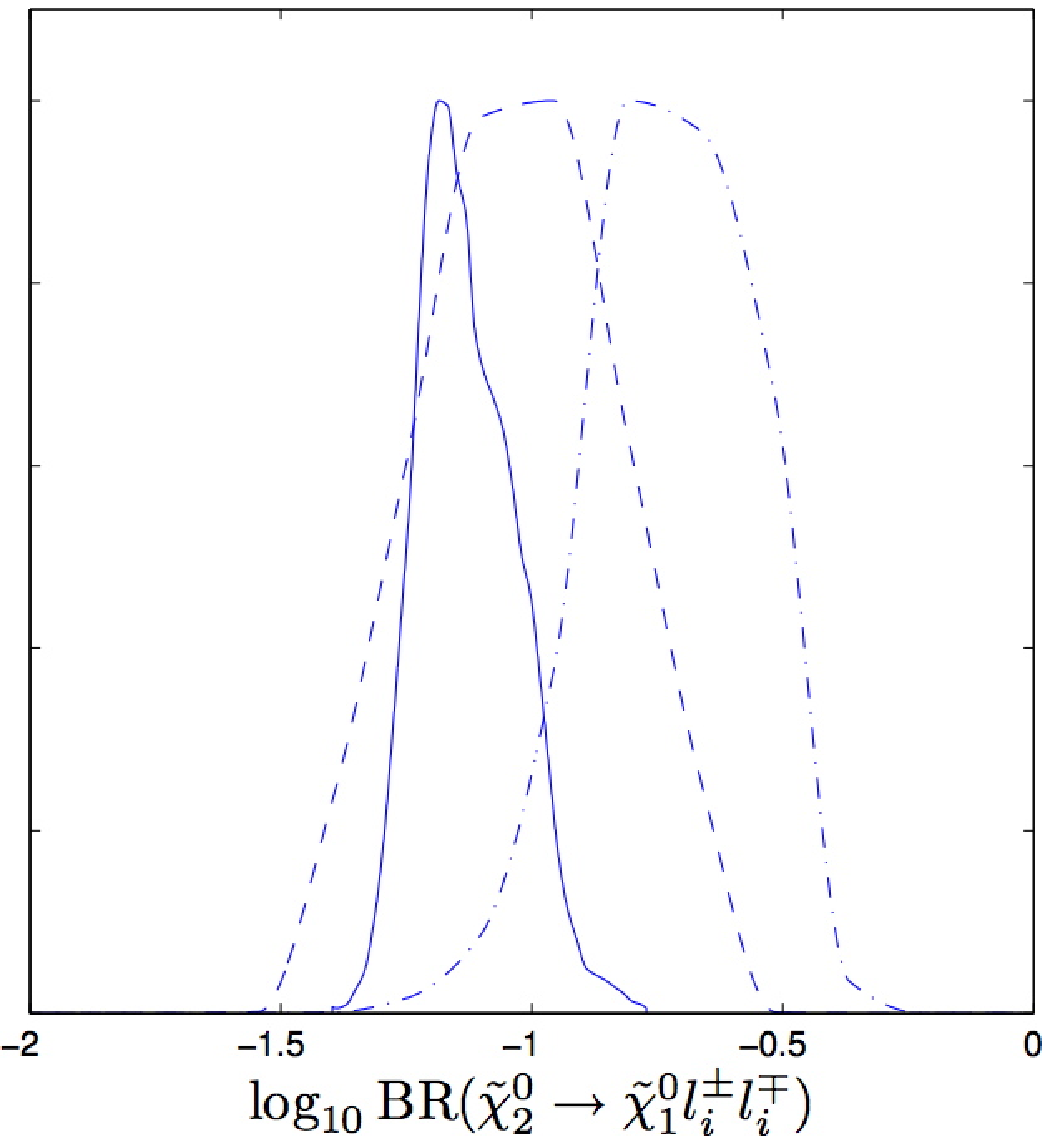}\quad
\includegraphics[height=5.5cm]{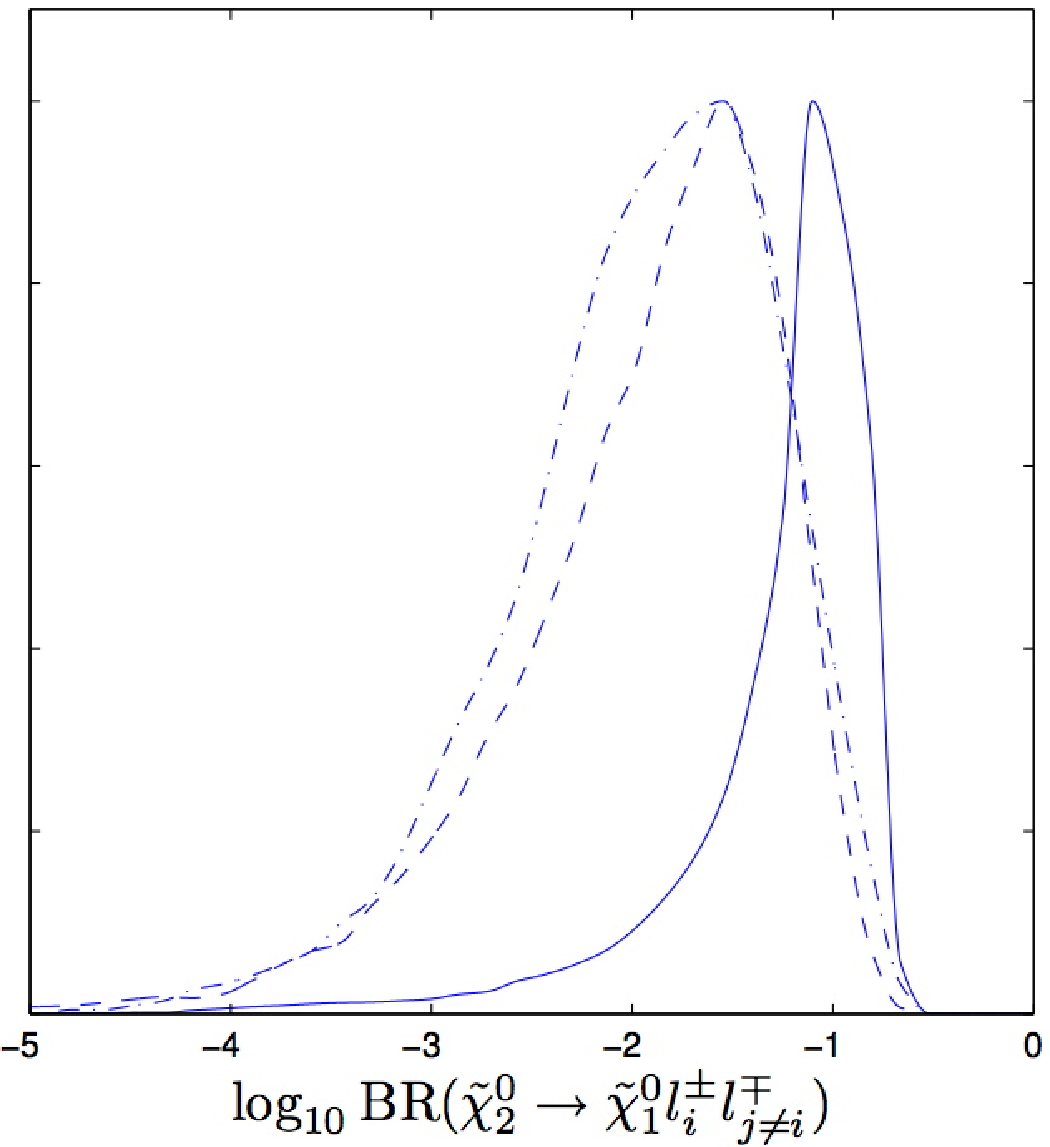}\quad
\includegraphics[height=5.5cm]{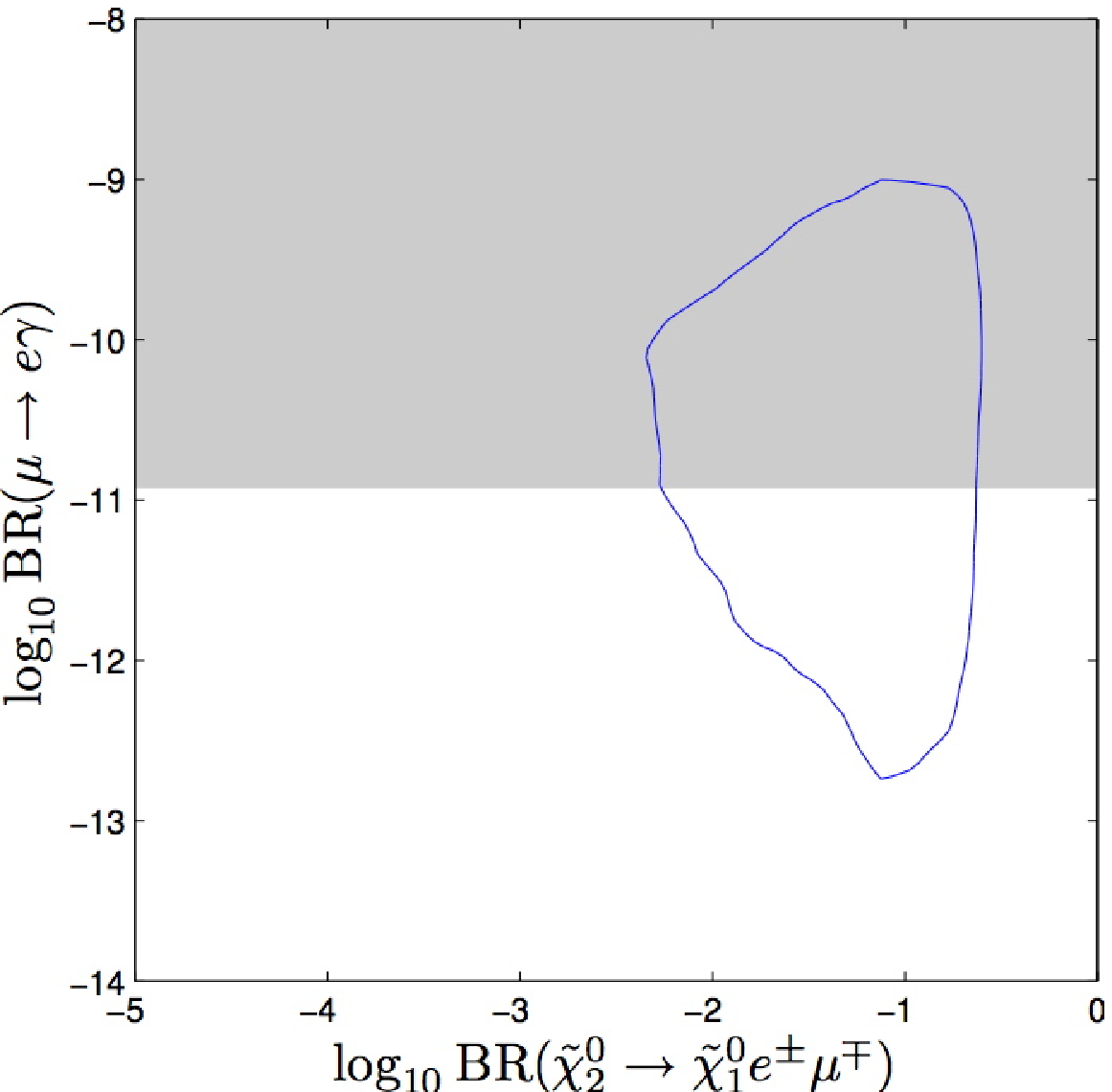}
  \caption{Probability densities for $\tilde\chi^0_2$ 
decays in a mixed brane-radion scenario, from \cite{Brummer:2011cp}.
The left and middle plots show  
$\textrm{BR}( \tilde{\chi}_2^0 \rightarrow l_i l_j \tilde{\chi}_1^0)$ 
with plain, dashed and dash-dotted lines corresponding to 
$l_il_i=ee$, $\mu\mu$, $\tau\tau$ (left) and 
$l_il_j=e\mu$, $e\tau$, $\mu\tau$ (middle), respectively.
On the right, $95$\% credible region in the 
$\textrm{BR}( \mu\rightarrow e\gamma)$ versus 
$\textrm{BR}( \tilde{\chi}_2^0 \rightarrow e \mu \tilde{\chi}_1^0 )$ plane. 
  }\label{fig:hgut}
\end{figure}

It was found that the most stringent constraints come from the 
bounds on lepton flavour violation, in particular from 
BR($\mu\to e\gamma$). The favourable regions of parameter space 
that satisfy the experimental constraints were examined with respect  to 
their LHC phenomenology. They generically feature heavy squarks 
and gluinos beyond the current LHC limit of 
$m_{\tilde q}\approx m_{\tilde g}\gtrsim 1.1$~TeV~\cite{lhc7}. 
The 5D setup leaves its imprints in the slepton mass matrices,  
which can lead to slepton mass splittings and 
interesting lepton-flavour violating SUSY decays 
at the LHC. The proliferation of unknown ${\cal O}(1)$ 
coefficients that occurs in this class of models necessitates  
a probabilistic approach to numerical predictions. 
As an example, Figure~\ref{fig:hgut} shows probability density distributions 
of $\tilde\chi^0_2\to l^\pm\tilde l^\mp \to l^\pm l^\mp\tilde\chi^0_1$ 
decays for a mixed brane-radion mediation scenario. 
The dilepton signature is often regarded as a gold-plated channel as 
it allows to obtain information on the neutralino and slepton masses 
from kinematic distributions. 
As can be seen, here dileptons with mixed flavours can have a sizeable rate. 
Extracting information from kinematic edges with flavour splitting and 
mixing has recently been studied in detail~\cite{Galon:2011wh}.
 
\begin{theacknowledgments}
TH thanks the organizers of the conference  for the interesting and 
valuable meeting  and the CERN theory group for its  hospitality during his regular visits to CERN where part of this work  was written.
\end{theacknowledgments}


\bibliographystyle{aipproc}

\end{document}